\documentclass{article}
\usepackage[utf8]{inputenc}
\usepackage{amsmath}
\usepackage{amssymb}
\usepackage{graphicx}
\usepackage{amsthm}
\usepackage{latexsym}
\usepackage{hyperref}

\newtheorem{lemma}{Lemma}
\usepackage{array}
\usepackage{tabularx}
\usepackage{makecell}

\usepackage{float}
\usepackage{algpseudocode}
\usepackage{algorithmicx}
\usepackage{algorithm}

\newcommand{\beq}{\begin{equation}}
\newcommand{\eeq}{\end{equation}}
\newcommand{\barr}{\left[\begin{array}}
\newcommand{\earr}{\end{array}\right]}

\newcommand{\bpf}{\begin{proof}}
\newcommand{\epf}{\end{proof}}

\newcommand{\ftwo}{\ensuremath{\mathbb{F}_{2}}}

\newcommand{\bi}{\begin{itemize}}
\newcommand{\ei}{\end{itemize}}
\newcommand{\bnum}{\begin{enumerate}}
\newcommand{\enum}{\end{enumerate}}
\newcommand{\bc}{\begin{center}}

\title{Black Box Cryptanalysis of AES128}
\author{Virendra Sule\thanks{Department of Computer Science and Engineering, Indian Institute of Technology Hyderabad, India. Email: viren.sule@gmail.com}
\and 
Kunal Telangi\thanks{
Private Researcher.
Email: kunaltelangi786@gmail.com}
}

\date{August 8, 2026}
\begin{document}
\maketitle
\begin{abstract}
This paper presents computational results of cryptanalysis of AES using the Local Inversion by Black Box computations of the forward encryption and utilizes these results to develop a practically feasible approach for the key recovery of the full scale AES128 under Known Plaintext Attack (KPA). It is shown that complete recovery of unknown key bits is possible upto $80$ bits in a practically feasible time and memory in random KPA situation by sequential computation when remaining $48$ bits are known. The results of key recovery in $64$, $72$ and $80$ bit unknown cases are extrapolated to predict the period of the iterative sequence generated in the local inversion approach for the full $128$ bit unknown key case and a strategy is proposed to search the actual period by brute force parallel search of the sequence period with $10$ free bits defining the search space. Then it is shown that the actual key can be verified in polynomial time by fast powering of the forward encryption map. Hence this strategy shows that the key recovery problem for AES128 under KPA has a chance of success in practically feasible time for a majority of the plaintexts. Local inversion approach to cryptanalysis using black box computations is a universal method applicable to a vast variety of key recovery and map inversion problems. Hence the results presented in this paper are representative of estimates of cryptanalysis of other ciphers which can be considered almost as strong as AES128 as encryption functions.   
\end{abstract}
Subject Classification: cs.CR, cs.CC, cs.SC, cs.LO, math.NT
\section{Introduction}
Advanced Encryption Standard (AES) is among the most widely deployed block cipher algorithms for bulk encryption in vast number of applications in internet and wireless communication as well as data storage. Since its standardization by NIST in 2001, the cipher has been cryptanalysed in several ways. It has remained resistant to especially algebraic attacks, linear and differential cryptanalysis in all its versions for last twenty five years. Only attacks which seem to have been successful in key recovery of AES are the side channel attacks. In this paper we formulate cryptanalysis of AES128 as the Local Inversion Problem (LIP) of maps using black box computations of the map. This approach has been termed as the Black Box Cryptanalysis (BBC) in previous papers \cite{Sule1,Sule2,Sule3} and is a universal approach to formulate cryptanalysis problems of a wide variety of cryptographic primitives. A brief description of BBC is given in the Appendix for details the reader is invited to look at the above references.

\subsection{Key recovery as local inversion of a map}
For a block encryption function $E(.,.)$ the ciphertext block $C$, the plaintext block $P$ and the key block $K$ are related by
\[
C=E(K,P)
\]
The Known Plaintext Attack (KPA) assumes that a pair $(P,C)$ of blocks is available and using the algorithm $E(.,.)$ and the data $(P,C)$ it is required to solve the key block $K$ and describe the computational requirement of solving for $K$. Define the function $F(x)=E(x,P)$. Then the key recovery under KPA is equivalent to solving the Local Inversion Problem (LIP)
\beq\label{LIP}
y=F(x)
\eeq
for given $y=C$. The computational approach to solving LIP (\ref{LIP}) is further constrained. We are only required to use the forward computation $E(x,P)$ for various values of $x$. Such operations are called black box operations. Number of such black box computations available for the solution of LIP is also limited by practical feasibility of sequential computation and memory. This approach to solving the LIP using black box operations is termed as Black Box Cryptanalysis (BBC).  

\subsection{BBC algorithm}
BBC shows that a much smaller set than brute force search is sufficient to solve for the key in certain special circumstances. Consider the sequence $S=\{Y(k)\}$ defined as follows when $y$ is given:
\begin{equation}\label{Sequence}
Y(k+1)=F(Y(k)), Y(0)=y
\end{equation}
for $k=0,1,2,\ldots$. If the sequence (\ref{Sequence}) is periodic there is $N$ such that 
\[
Y(N+j)=Y(j)
\]
for $j=0,1,2,\ldots$. Hence the finite sequence $\{Y(k),k=0,1,2,\ldots,(N-1)\}$ satisfies $F(Y(N-1))=y(0)$. Hence one solution of the key is the last element of the periodic sequence. The sequence generation uses only black box operations. The solution to the inverse
\[
x=Y(N-1)
\]
is called the \emph{orbit inverse}. Another case of the sequence $S$ that can occur is the quasi-periodic case for which there exist $T$ the tail length (or quasi period) and and period $N$ such that $Y(T+j)=Y(N+T+j)$. For such a sequence a random permutation $\phi$ of bits of $Y(0)$ and $F(.)$ can lead to a periodic sequence whose orbit inverse leads to the solution of the original LIP for $Y(0)=y$. This procedure is described in Appendix. Consider the algorithm:
\subsubsection{BBC Algorithm}\label{BBC}
\begin{enumerate}
    \item Input: $y$ for the local inversion problem.
    \item Generate sequence $S=\{Y(k),k=0,1,2,\ldots\}$ upto a finite maximum length $M$. 
    \item Find if the sequence is periodic of period $N\leq M$.
    \item One possible solution to the key is the orbit inverse $Y(N-1)$.
    \item If the sequence is quasi-periodic, generate a periodic sequence by using a random permutation of $Y(0)$ and solve for the orbit inverse to get the solution of the original LIP. Details of the this process are given in the Appendix.
\end{enumerate}
This is called the BBC algorithm at a primary level and can compute the orbit inverse as a solution. The largest value of $M$ is the limit to feasible search which can be fixed based on the limit of feasible time. The memory required is the storage of one step and one comparison. This approach is reminiscent of the Time Memory Tradeoff (TMTO) but requires much smaller storage. It is also well known from the analysis of the Brent algorithm \cite{Cohen} that the period $N$ of a periodic cycle or $T+N$ (tail length $T$, period $N$) of a quasi periodic cycle are both detected with at least $50$\% chance in $O(2^{n/2}$ and have values of $T+N$ approximately $1.25*2^{n/2}$. Hence the algorithm has a chance of being practically feasible if the periods of the cycles are small enough to be practically feasible for sequential computation. Cycles of length upto $2^{38}$ are considered to be practically feasible. 

This approach can however still fail in the following case. There is a nonunique solution to the LIP. There exists $\tilde{x}$ such that $F(\tilde{x})=y$ hence $\tilde{x}$ cannot be captured by the sequence $S$ as an orbit inverse. This may happen because the function $F$ is designed to be highly non-linear. The practical experience obtained from the AES case studies that are reported in this paper shows that this case of nonunique inverse never appeared.

The BBC algorithm can thus expect to compute the orbit inverse with the sequential search over $N$ which may also be of exponential order $O(2^{n/2})$ while brute force search requires search over space of order $2^n$ where $n$ is the key length. The case studies reported in this paper showed that the actual period $N$ always fell below the square root order $O(2^{n/2})$ and was higher than the sub-exponential order $O(2^{4*\sqrt{n}})$ for various values of $n$.

\subsubsection{Logic of strategy for Key recovery in $128$ bit case}
The above strategy of sequentially computing the period of the sequence (\ref{Sequence}) to find the key as $Y(N-1)$ is not computationally feasible in $128$ bit case because the period is too large to be computed sequentially on commonly available computers. The period computation algorithms require computations of the order $O(2^{64})$. However, the random trials in lower bit cases $64,72,80$ show that there is a clustering of the periods of sequences (\ref{Sequence}) in these lower bit cases within a sharp band of a maximum period. Hence if the maximum observed periods of lower bit cases are extrapolated for $128$ bit case, then it is reasonable to expect that majority of periods of sequences (\ref{Sequence}) over plaintext space will be near a sharp band around an expected maxium period. Using this logic we develop a strategy to solve the key recovery problem in the $128$ bit case by brute force searching  of the period for given $P$ over a small range of $10$ free bits by parallel computation. This attack is described in Section 3 and has a potential to solve the key recovery problem in $128$ bit case in practically feasible time for a majority of cases of known plaintexts. Analysis of actual percentage of predicted success of these cases shall be explored in a subsequent paper.        

Thus the local inversion approach for cryptanalysis is purely an empirical approach and does not depend on the complexity of construction of the AES functions. The approach purely depends on black box computations, fast powering of maps by black box operations as well permutations of $n$ bits instead of that of the space of $2^n$ points. These aspects make this approach practically feasible.

\section{Local Inversion for Partial Key Recovery}
In this section we show how the BBC algorithm \ref{BBC} can be utilized for recovering partially unknown key bits of the full $128$-bits of the key of AES. Such a procedure facilitates a uniformly structured small key length cipher models constructed from the standard AES128. Hence actual construction of analogous building blocks for a reduced scale cipher resembling standard AES128 is not necessary. We shall call these as \emph{partially known key models} of AES128.

The central idea of partially known key model and its key recovery is described as follows. Let 
\[
M=(m_0,m_1,\ldots,m_{127})
\]
denote the array of $128$ bits of the master key of the standard AES128. Next, consider a notation to describe partial bits of vectors. Let $X$ denote an arbitrary array of $q$ bits and $Q$ denote the ordered set of all indices of bits of $X$,
\[
Q=\{1,2,\ldots,q\}
\]
for a subset $U\subset Q$ of indices, let $X|U$ denote the subarray of $X$ with bits of indices in $U$ in the same order as listed in $U$ while by $X|U'$ to be the subarray of $X$ of indices in $Q$ in the complement of $U$ but in the same order as in $Q$. 

\subsection{LIP for recovering partial key bits}
Using these notations we now describe the LIP (\ref{LIP}) for recovering a specific subset of unknown bits of the master key $M$ of the original LIP 
\[
y_0=F(M)
\]
In this problem we are given a partial set of indices $U$ in the set of all $n=128$ indices of $M$. We then pose the problem of solving for the subarray $x|U$ where the subarray $x|U'$ is already fixed or given.

\subsection{Iterative sequence for partial key recovery}
As explained in the Introduction, the BBC approach to solve the LIP seeks the orbit inverse of an iterative sequence generated by the map of the above LIP. Let $y_0$, the output of the map and the map itself be defined by one plaintext ciphertext pair $(P,C)$. Then define the iterative sequence as follows: (Let $M$ denote the Master key. It is assumed that the partial array $M|U$ is unknown while the remaining array bits $M|U'$ are fixed and known).
\beq\label{Seqpartialkey}
Y(k+1)=F(X(k))
\eeq
where $Y(0)=y_0$ and $X(k)|U=Y(k)|U$, $X(k)|U'=M|U'$. Compactly we may also define the new map by assuming the known bits $M|U'$ to be
\[
Y(k+1)=\tilde{F}(Y(k))=F(\phi(Y(k))
\]
where $Y(0)=y_0$ and $\phi(Y(k))|U=Y(k)|U$, $\phi(Y(k))|U'=M|U'$.

Algorithms for key recovery are then aimed at recovering the unknown part $M|U$ as an orbit inverse of the sequence $Y(k)$. 

\subsubsection{BBC Algorithm for partial key recovery for feasible period}\label{BBCpartial}
The cases in which the sequence is periodic can be handled by the following steps:
\begin{enumerate}
    \item The sequence is periodic of period $N$. Then $Y(N)=Y_0$. Hence by the iteration rule $Y(0)=F(\phi(Y(N-1))$. Hence $\phi(Y(N-1))|U=Y(N-1)|U=M|U$ is the unknown key. 
    \item Hence in actual computation the simplest way to discover the key is to compute the period of the sequence (\ref{Seqpartialkey}) and storing the previous value of the term in the sequence to find $Y(N-1)|U$. If the period $N$ is of feasible order in $n$ the computation is successful in feasible time.
\end{enumerate}
The conceptual flow chart of the algorithm is described in the following figure.

\begin{figure}[H]
\centering
\includegraphics[width=0.6\textwidth]{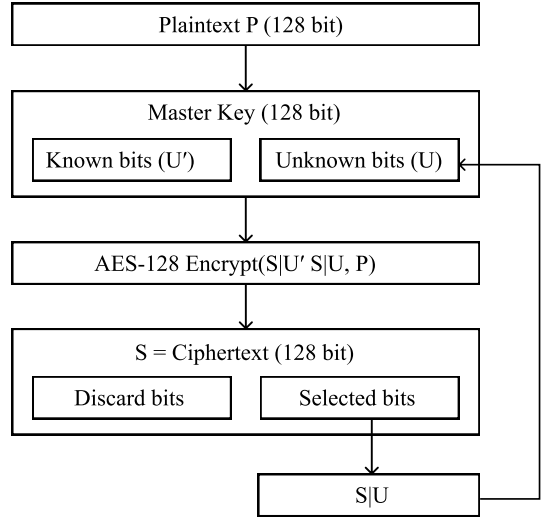}
\caption{Iterative state generation for the AES128 LIP under partially known key bits.}
\label{fig:aes}
\end{figure}

The local inversion is thus carried out by a period finding algorithm on the sequence $S$. Hence this computation is possible sequentially only when the period $N$ is of feasible order in $n$.

\subsection{Results for partial key recovery of $64,72,80$ bits in AES128}
Results shown in Table \ref{tab:aes_64_64} are for 
128-bit AES cycle detection and key recovery of 64-bit known master key part ($M_0, M_1$) the first four $64$ bits and 64-bit unknown master key part ($M_2, M_3$) the last four $64$ bits. All experiments were performed on a processor operating at a base clock speed of 4.3 GHz and a maximum boost clock speed of up to 5.7 GHz, with 32 GB RAM.

\begin{figure}[H]
\centering
\includegraphics[width=1\textwidth]{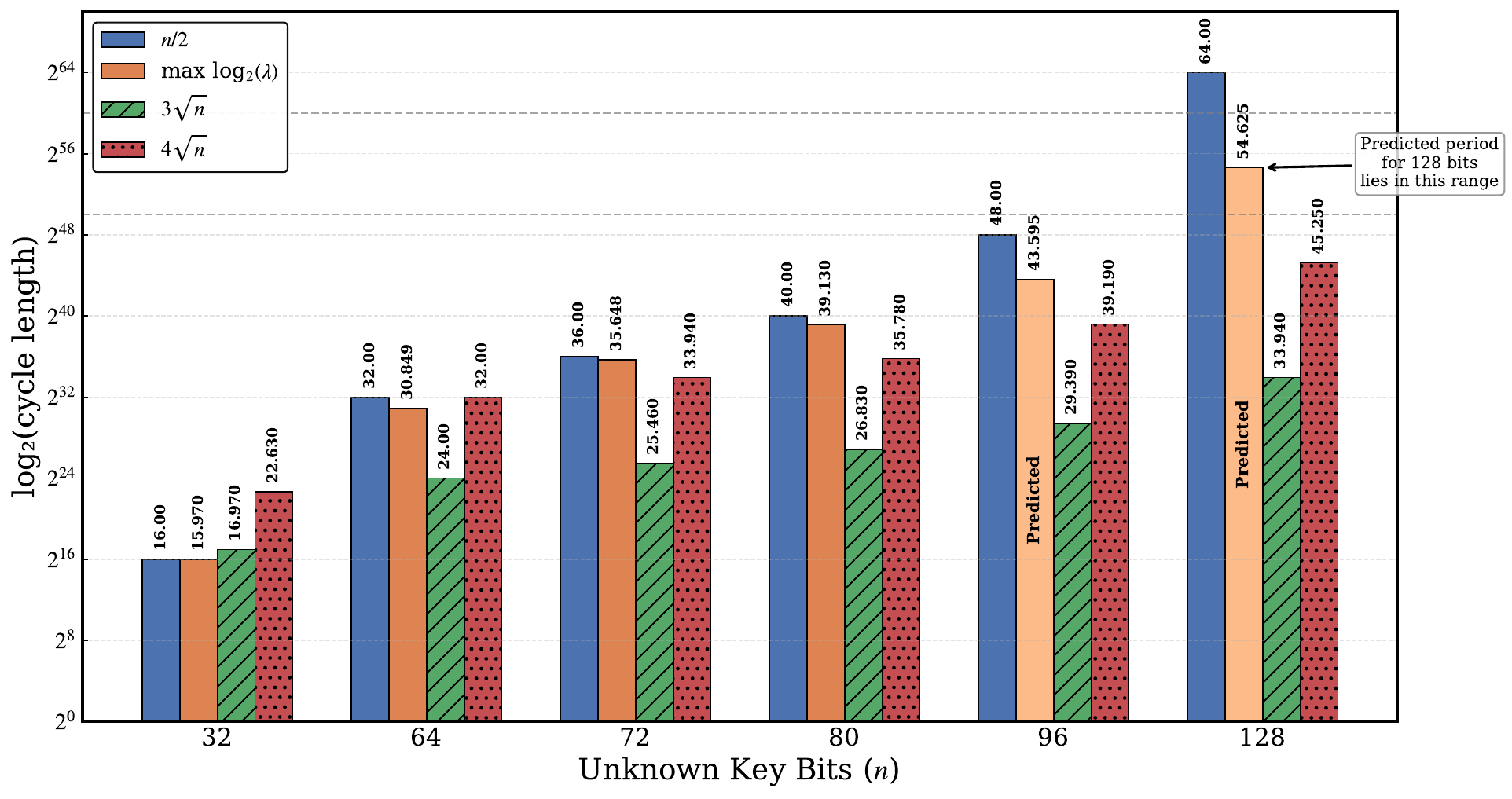}
\caption{Bar chart comparing the observed trends of $\log_2(\lambda)$ with respect to the number of unknown key bits $n$ for the full-round AES-128 LIP, where $\lambda$ denotes the cycle length.}
\label{fig:aes_period_comparison}
\end{figure}

From the comparison of the tables 1 and 2 it follows that the periods of the output sequences (cycle lengths) appear to follow a sub-exponential to exponential order pattern in the number of unknown bits of the key in the range $2^{4\times\sqrt{72}}=2^{34}$ to $2^{72/2}=2^{36}$. Hence the time taken to recover the key from the last element of the periodic sequence is also in a short range of time $2-12$ minutes. Hence whether the unknown key bits are random or in a continuous run within the master key has caused no drastic variation. For the $80$ bit unknown case also the period seems to be bounded by the sub-exponential to exponential pattern $2^{4*\sqrt{64}}=2^{32}$ to $2^{40}$. Hence following the period pattern upto $80$ bit unknowns the period for the LIP of full $128$ bit unknown key is expected to be bounded between $2^{5*\sqrt{128}}=2^{56}$ to $2^{64}$. This predicted period is shown in figure \ref{fig:aes_period_comparison}. 

15 Independent random experiments were conducted under the 64 bit known and 64 bit unknown master key split. Of these, 11 experiments successfully recovered the master key. The remaining 4 experiment resulted in a quasi cyclic.

\begin{table}[H]
\centering
\caption{Computations for recovering unknown 64 bits of Master key}
\label{tab:aes_64_64}
\small
\renewcommand{\arraystretch}{1.3}
\setlength{\tabcolsep}{4pt}
\begin{tabularx}{\textwidth}{|>{\ttfamily\scriptsize\centering\arraybackslash}X|
>{\ttfamily\scriptsize\centering\arraybackslash}X|
c|c|c|}
\hline
\makecell{\textbf{Plaintext}\\\textbf{(P)}} &
\makecell{\textbf{Master Key}\\\textbf{(MK)}} &
\makecell{\textbf{Cycle}\\\textbf{Length ($\boldsymbol{\lambda}$)}} &
\makecell{\textbf{Time}\\\textbf{in min}} &
\makecell{\textbf{Key}\\\textbf{Recovered}}
\\
\hline
\makecell{70b66c70\\bc7bbbf7\\20a9a2b1\\fe66d71b} &
\makecell{dc779966\\7dff3590\\50d602c3\\ddb46e19} &
$2^{23.687}$ & 0.01 min & Yes \\
\hline
\makecell{a034696f\\b085673c\\d30a4d38\\39957660} &
\makecell{b78d5bdb\\d66bc420\\fae1e2c5\\be70b2bd} &
$2^{28.815}$ & 0.27 min & Yes \\
\hline
\makecell{024593f3\\f85c2e92\\8f52fe4d\\07432442} &
\makecell{cf3142c9\\af8d850d\\b82ab9e0\\bd7aee5d} &
$2^{27.080}$ & 0.08 min & Yes \\
\hline
\makecell{5b18cd0c\\4390626d\\b4fd4031\\c359ff07} &
\makecell{8177f41e\\a014f6e8\\8a29ead0\\169f5147} &
$2^{29.754}$ & 0.51 min & Yes \\
\hline
\makecell{c10cb361\\88d63d77\\143e2748\\e9c5e7df} &
\makecell{dba4d67d\\8cb340d0\\a0dca41a\\646e5348} &
$2^{30.686}$ & 0.97 min & Yes \\
\hline
\makecell{d48592b4\\58bc1822\\e7dbea8c\\551e043d} &
\makecell{903755d1\\a78b15cc\\30c23917\\333a3592} &
$2^{30.849}$ & 1.09 min & Yes \\
\hline
\makecell{cfde424e\\c233973b\\54fd7020\\21175e3f} &
\makecell{29971aa5\\d6a47c10\\6d2ba743\\2dc18e9c} &
$2^{28.610}$ & 0.23 min & Yes \\
\hline
\makecell{9df44143\\1f7a2768\\dcab4733\\5919839a} &
\makecell{ddcc89fb\\fcdcb147\\fc0b67bb\\8198fe8c} &
$2^{23.577}$ & 0.01 min & Yes \\
\hline
\makecell{2eb4fe2c\\51194dc6\\ac9bdc15\\222c5367} &
\makecell{2ee8dc6e\\9d9a2464\\9624860e\\22b96c05} &
$2^{28.645}$ & 0.24 min & Yes \\
\hline
\makecell{1fd3251d\\7b58e080\\402f2933\\55d008e7} &
\makecell{d6424e26\\7dbbc3b2\\047bd756\\80cca4fb} &
$2^{29.763}$ & 0.51 min & Yes \\
\hline
\makecell{8339476d\\f322657e\\ffa55b63\\b51cccc2} &
\makecell{d810c62a\\ac39eef1\\299f3d6d\\5475f3d6} &
$2^{27.216}$ & 0.09 min & Yes \\
\hline
\end{tabularx}
\end{table}

Table \ref{tab:aes_64_64} shows results for recovery of random $64$ bits of the master key of AES while the remaining bits are held constant. The known bit positions (0-based, from LSB to MSB) are held constant and equal to the corresponding bits of the initial random key; the remaining 64 bits are taken from the current ciphertext.

\begin{table}[H]
\centering
\caption{Computations for recovering unknown 72 bits of Master key (56 known bits)}
\label{tab:aes_72_56}
\small
\renewcommand{\arraystretch}{1.3}
\setlength{\tabcolsep}{4pt}
\begin{tabularx}{\textwidth}{|>{\ttfamily\scriptsize\centering\arraybackslash}X|
>{\ttfamily\scriptsize\centering\arraybackslash}X|
c|c|c|}
\hline
\makecell{\textbf{Plaintext}\\\textbf{(P)}} &
\makecell{\textbf{Master Key}\\\textbf{(MK)}} &
\makecell{\textbf{Cycle}\\\textbf{Length ($\boldsymbol{\lambda}$)}} &
\makecell{\textbf{Time}\\\textbf{in min}} &
\makecell{\textbf{Key}\\\textbf{Recovered}}
\\
\hline
\makecell{4df17f3c\\bd911104\\c063e967\\f5ddc21d} &
\makecell{00446d8d\\c5f927c4\\62ba217e\\d96622bf} &
$2^{33.644}$ & 7.51 min & Yes \\
\hline
\makecell{35326178\\db6fe364\\e882ac99\\cff11b54} &
\makecell{f93b9f8a\\f7b5178b\\38ff4ccc\\d99b513d} &
$2^{34.011}$ & 9.68 min & Yes \\
\hline
\makecell{e41a2b5a\\b4e4f4ba\\33e11586\\c901985c} &
\makecell{4fa5ce45\\0d3a608b\\84ff9032\\d523aa0f} &
$2^{35.363}$ & 24.73 min & Yes \\
\hline
\makecell{ff478605\\977c4110\\044ee265\\8d45ae0b} &
\makecell{a463bc6e\\390ba124\\07aad993\\7291e2ac} &
$2^{34.869}$ & 17.55 min & Yes \\
\hline
\makecell{7ce2adf9\\e5aaac2c\\af0c6d38\\4bc6bea1} &
\makecell{441bc07b\\70b416d1\\b1163fc9\\48d50180} &
$2^{34.848}$ & 17.30 min & Yes \\
\hline
\makecell{d92a725e\\acf45ad7\\51ebd1c5\\d6e3dd33} &
\makecell{f9056b19\\fc3e6b9f\\6ff3edee\\ad8c7a14} &
$2^{35.648}$ & 30.11 min & Yes \\
\hline
\end{tabularx}
\end{table}
\begin{table}[H]
\centering
\caption{Computations for recovering unknown 80 bits of Master key (48 known bits)}
\label{tab:aes_80_48}
\small
\renewcommand{\arraystretch}{1.3}
\setlength{\tabcolsep}{4pt}
\begin{tabularx}{\textwidth}{|>{\ttfamily\scriptsize\centering\arraybackslash}X|
>{\ttfamily\scriptsize\centering\arraybackslash}X|
c|c|c|}
\hline
\makecell{\textbf{Plaintext}\\\textbf{(P)}} &
\makecell{\textbf{Master Key}\\\textbf{(entry)}} &
\makecell{\textbf{Cycle}\\\textbf{Length ($\boldsymbol{\lambda}$)}} &
\makecell{\textbf{Time}\\\textbf{in min}} &
\makecell{\textbf{Key}\\\textbf{Recovered}}
\\
\hline
\makecell{37f2c031\\4002fdc2\\4c965e69\\d5e1109b} &
\makecell{74c8d4d3\\bcfe9a0e\\581f6b8e\\5579fcfd} &
$2^{38.98}$ & 284.78 min & Yes \\
\hline
\makecell{fee634d7\\bcc139fa\\9f359e49\\6f6644c5} &
\makecell{5a31feaa\\4caf08a6\\e5d45642\\f8cabbf2} &
$2^{37.16}$ & 84.06 min & Yes \\
\hline
\makecell{eb7bb3e1\\b4d3e2da\\3ee22b10\\44669cac} &
\makecell{b414e65f\\469aa26b\\ba73d9cb\\ba1e7088} &
$2^{37.605}$ & 114.46 min & Yes \\
\hline
\makecell{32af77b4\\28818549\\85331d2e\\c3cfc9d5} &
\makecell{08ea54b9\\c4771155\\c67e6ce9\\5741a9f7} &
$2^{39.13}$ & 329.01 min & Yes \\
\hline
\makecell{4ba3a9d5\\1cdbb5a5\\90cccccb\\2353f4ab} &
\makecell{3205c32a\\7c9d64f5\\05bc7a7b\\35234f62} &
$2^{37.37}$ & 97.28 min & Yes \\
\hline
\end{tabularx}
\end{table}

An important observation about these computations is worth stating. When $U$ denotes a subset of key bits denoting the partical key $K|U$, it is found that the actual location of $U$ in the full set of $128$ bits does not have an appreciable impact on the period lengths of sequences. Periods of iterative sequences for different subsets $U$ of same cardinality were approximately close to each other. This may be due to the balanced-ness of the pseudorandom behaviour of the forward map.

\section{Full key recovery of AES128}
In this section we discuss the strategy to do the full $128$ bit key recovery for the given plaintext ciphertext pair $(P,C)$ of AES128. This strategy consists of following two steps
\begin{enumerate}
    \item Guess a range where the possible period of the iterative sequence $Y(k+1)=F(Y(k))$ for $Y(0)=y=C$ and $F(x)=E(x,P)$ belongs for majority of cases of KPA. Determine the actual period $N$ by verifying the equation $Y(0)=F^{(N)}(Y(0))$ in parallel search. The computation of $F^{(N)}(Y(0))$ for each of these searches can be computed in polynomial time by fast powering of the black box computation of the forward map.
    \item Once $N$ is verified, compute $x=Y(N-1)$.
\end{enumerate}
 Thus it follows that if we can find a short range for the possible key the number of parallel searches is correspondingly smaller in majority of the cases of KPA. Hence it is required to fix the upper bound on feasible search space for the possible periods.

 \subsection{Predicted possible periods for $128$ bit unknown key}
 The plots of maximum periods of random samples of $(P,C)$ periods are shown in figure \ref{fig:aes_period_comparison} for $64$, $72$ and $80$ bits. All these actual periods belong to the range from sub-exponential order $2^{5*\sqrt{n}}$ to the exponential order $2^{n/2}$. For $128$ bits this range is between $2^{56}$ to $2^{64}$. Hence this involves $8$ bits of free choices of periods. Extending this range to $10$ bits to consider periods from $2^{54}$ to $2^{64}$ seems to be highly probable for an actual period to belong. Hence we consider the periods to be searched over $10$ bits in the general expression
\[
N=\sum_{i=0}^{i=53}2^i+\sum_{i=54}^{i=63}a_i2^i
\]
for arbitrary bits $a_i$. This is a collection $\Omega$ of $1024$ numbers for possible periods of the sequence $S$ for full $128$ bit unknown key for a given $(P,C)$ pair. Then execute the following algorithm

\subsubsection{Algorithm: Search for actual period and key recovery}
\begin{enumerate}
    \item Compute all $N$ in the above collection $\Omega$ fixed by $10$ bit strings of $a_i$.
    \item Fix a set of random permutations of $n$ bits of points in $\ftwo^n$.
    \item Develop a bank of squarings of maps $F$ and $\tilde{F}=\phi\circ F$ upto largest squaring needed to compute fast powering of $F^(N)(.)$, $\tilde{F}^(N)(.)$ at any input.
    \item For given $Y(0)$ compute $\tilde{Y}(0)=\phi(Y)(0)$.
    \item For each of the numbers $N$ in the collection $\Omega$ compute $F^{(N)}(Y(0))$, $\tilde{F}(\tilde{Y}(0))$ by fast powering. Then verify whether $Y(0)=F^{(N)}$ or $\tilde{Y}(0)=\tilde{F}^(N)(Y(0))$. Return the $N$ at which one of the expressions is satisfied.
    \item Compute $N-1$ and the key $K=F^{(N-1)}(Y(0))$ or $K=\tilde{Y}(N-1)$.
\end{enumerate}

\subsubsection{Fast powering of maps is feasible}
Fast powering of computing $F^{(N)}(Y(0))$ is obtained by preparing computer codes for squared maps $F^{(2^i)}$. Where, if a computer code or a function $FunctionF()$ represents a map $F$ then the map $F^{(2)}$ is represented by the function
\begin{verbatim}
FunctionF2()
Input: x
y ← FunctionF(x)
y ← FunctionF(y)
Return output y
\end{verbatim}
as the code corresponding to the squared map. Similarly $F^{(2^i)}$ is represented by the squaring the code of the map $F^{(2^{i-1})}$. We can now state the following fact which is easy to prove.

\begin{lemma}
    Let $F(.)$ is a map whose code $FunctionF()$ executes in polynomial time for arbitrary input $x$, then the code of the squared map $F^{(2)}$ described above also executes in polynomial time for all inputs.  
\end{lemma}

Next, it follows that the map power $F^{\sum_{i=m}^{i=n} a_i2^i}(x)$ at any input $x$ is the composition 
\[
(F^{(2^m)})^{a_m}\circ\ldots\circ(F^{(2^n)})^{(a_n)}(x)
\]
which is also represented by a code which executes in polynomial time. This proves that the map powering $F^{(N)}(Y(0))$ can be computed in polynomial time. This is termed as fast powering of a map evaluation at a given input. 
\section{Appendix: Brief description of BBC}
In this approach to cryptanalysis the key recovery problem of an encryption function 
\[
C=E(K,P)
\]
under the Known Pliantext Attack (KPA) is considered as the Local Inversion Problem (LIP) of a map $F:\ftwo^n\rightarrow\ftwo^n$. The LIP is concerned with solving $x$ in $\ftwo^n$ for a given $y$ in $\ftwo^n$ such that
\beq\label{LIP}
y=F(x)
\eeq
The main constraints considered for solving the LIP using the BBC approach are
\begin{enumerate}
    \item The computation should only utilize the data obtained by black box (or forward) computation of the map $F$. In the KPA situation, this is the forward computation of $C$ given $P$ for any input $K$.
    \item The number of forward computations should be practically feasible. (or of order $O(P(n))$ in number of bits of the key where $P(n)$ has practically feasible value for the actual key length $l$. 
\end{enumerate}
Hence the BBC approach restricts the number of computations and memory required for solving the LIP to a pre-defined feasible order. Within such a restriction, the aim of BBC is to determine the estimate of number of successful cases of key recovery for the algorithm empirically. 

\subsection{Solving LIP as inversion of a periodic sequence}
The strategy of the BBC is to formulate the LIP as a problem of sequence inversion of the iterative sequence
\beq\label{Seq}
S=\{y(k),k=0,1,2,\ldots,(M-1)\}
\eeq
defined as $y(0)=y$ where $y$ is given data on LIP and
\[
y(k+1)=F(y(k))
\]
The length $M$ of the sequence is fixed to be practically feasible and of order $O(P(n))$ in $n$ the bit length. When $S$ is periodic of period $N$, the \emph{Orbit inverse} $y(-1)$ is the term $y(N-1)$. There are likely to be cases when the sequence $S$ is not periodic but quasi-periodic hence there exist the tail length $T$ and period $N$ such that $y(T)=y(T+N)$. For quasi periodic sequences the local inversion $y(-1)$ for the problem $y(0)=F(x)$ is obtained by using a random perputation of bits of $y(0)$ and the outputs of the map $F$ as follows.

\subsubsection{Sequence Inverse for quasi-periodic case}
Consider the sequence $S$ generated by $F$ and initial value $y=Y(0)$ to be quasi-periodic. A procedure to convert a quasi-periodic cycle into a cycle which high chance of succeeding is well known and was given by Hellman \cite{Hellman}. Consider a random permutation $\phi$ of $\{0,1\}^n$. The LIP is then
\[
\tilde{Y}(0)=\phi(Y(0))=\phi(F(x))=\tilde{F}(x)
\]
If the sequence $\tilde{S}$ generated by the iteration
\[
\tilde{y}(k+1)=\tilde{F}(\tilde{y}(k))
\]
is periodic of period $\tilde{N}$, then we get the orbit inverse of $\tilde{S}$ as 
\[
\tilde{Y}(\tilde{N}-1)=\tilde{F}^{(\tilde{N}-1)}=\tilde{F}^{(\tilde{N}-1)}(\tilde{Y}(0)) 
\]
Since
\[
\tilde{Y}(0)=\tilde{Y}(\tilde{N})=\tilde{F}(\tilde{F}^{(\tilde{N}-1)}(Y(0))
\]
it follows that after inverting $\phi$ from both sides of above equation we get
\[
Y(0)=F(\tilde{F}^{(\tilde{N}-1)}(\tilde{Y}(0))
\]
Hence
\[
x=\tilde{F}^{(\tilde{N}-1)}(\tilde{Y}(0))=\tilde{Y}(N-1)
\]
is the actual inverse of the original LIP. This strategy of random permutation to solve the local inversion is likely to work for high pseudorandom sequence generating maps $F$. If one $\phi$ fails to produce a periodic sequence then another random $\phi$ may be selected. 

The above description shows that periods of sequences generated by $F$ and initial $y$
can be computed by cycle detection algorithms such as Pollard rho or Floyed cycling which require sequential time of the order of the period and only one storage of a sequence element. Hence when the periods are of practically feasible order of complexity in the number of bits $n$ the local inversion can be computed in feasible time. Well known analysis of the period finding algorithm for sufficiently random maps $F$ \cite{Cohen} shows that periods $N$ or the values of $T+N$ for quasi periodic sequences have a $50$\% chance of being detected in $O(2^{n/2})$ and also have value approximately $1.25*2^{n/2}$. Hence the above procedures of cycle detection and conversion of quasi cycles into cycles by permutation have a chance of being practically feasible if the actual periods of the iterative sequences generated by $F$ and initial value $Y(0)$ are small enough for sequential computation.  

\section{Conclusions}
This paper provides several empirical results about key recovery of AES128 for smaller number of unknown key bits $64$, $72$ and $80$ in random KPA situation in practically feasible time and memory by sequential computation. These results were feasible primarily because the periods of iterative sequences of the LIP were of small order than even $O(2^{n/2})$ for $n$ bit unknowns upto $80$ bits. These results led to the conclusion that the period for full $128$ bit unknown key in a random KPA can be predicted to be within a small band of number of unknown bits such as $10$ bits to search for the period. Once an actual period is verified in parallel for each of the numbers in the search, the key can be recovered in practically feasible time due to the high efficiency of the black box forward operation of AES and fast exponential powering of the forward map. This leads to the striking conclusion that AES128 has a high chance of being broken with complete key recovery in a single pair KPA in practically feasible time and memory for a majority of plaintexts. The black box approach to local inversion proposed in this paper only utilizes the input output data of the map to be inverted. Hence it is expected that empirical results of AES128 bit are likely to be applicable approximately to several other $128$ bit unknown encryption functions.


\begin{thebibliography}{99}
    \bibitem{Cohen} H.\ Cohen. A Course in Computational Algebraic Number Theory. Vol.138, Graduate Texts in Mathematics, Springer, Berlin, 1993.
    \bibitem{Hellman} Martin Hellman. A cryptanalytic time-memory trade-off. IEEE Trans. on Information Theory, 26(4), pp.401-406, 1980.
    \bibitem{StampLow} Mark Stamp and Richard M.\ Low. Applied Cryptanalysis. Wiley-Interscience, 2007.
    \bibitem{Sule3} Virendra Sule: Local inversion of maps: Black Box Cryptanalysis. \texttt{http://arXiv.org/cs.CR/2207.03247v2. July 2022}.
\bibitem{Sule2} Virendra Sule. Local Inversion of Maps: A New attack on Symmetric Encryption, RSA and ECDLP.
\texttt{http://arXiv.org/cs.CR/2202.06584v2. March 2022}.
\bibitem{Sule1} Virendra Sule: A complete algorithm for local inversion of maps: Application to Cryptanalysis.
\texttt{http:\\arXix.org/cs.CR/2202.06584v2. May 2021}.
\end{thebibliography}
\end{document}